\documentclass{emulateapj}
\usepackage{apjfonts}

\usepackage{graphicx,natbib}

\begin{document}

\title{Simulations of AGN feedback in galaxy clusters and groups: impact on
  gas fractions and the $L_{\rm X}-T$ scaling relation}

\shorttitle{Simulations of AGN feedback in galaxy clusters and groups}
\shortauthors{E. Puchwein, D. Sijacki and V. Springel}

\author{E. Puchwein\altaffilmark{1},
        D. Sijacki\altaffilmark{2},
        V. Springel\altaffilmark{1}}

\affil{$^{1}$Max-Planck-Institut f\"{u}r Astrophysik,
      Karl-Schwarzschild-Stra\ss{}e 1, 85740 Garching bei
      M\"{u}nchen, Germany}
\affil{$^{2}$Institute of Astronomy, Madingley
      Road, Cambridge, CB3 0HA, United Kingdom}


\email{puchwein@mpa-garching.mpg.de}

\begin{abstract}
  Recently, rapid observational and theoretical progress has established that
  black holes (BHs) play a decisive role in the formation and evolution of
  individual galaxies as well as galaxy groups and clusters. In particular,
  there is compelling evidence that BHs vigorously interact with their
  surroundings in the central regions of galaxy clusters, indicating that any
  realistic model of cluster formation needs to account for these
  processes. This is also suggested by the failure of previous
  generations of hydrodynamical simulations without BH physics to
  simultaneously account for the paucity of strong cooling flows in clusters,
  the slope and amplitude of the observed cluster scaling relations, and the
  high-luminosity cut-off of central cluster galaxies. Here we use
  high-resolution cosmological simulations of a large
  cluster and group sample to study how BHs affect their host systems. We focus on
  two specific properties, the halo gas fraction and the X--ray luminosity-temperature scaling relation, both of which are notoriously difficult to
  reproduce in self-consistent hydrodynamical simulations. We show that BH
  feedback can solve both of these issues, bringing them in
  excellent agreement with observations, without alluding to the `cooling
  only' solution that produces unphysically bright central galaxies. By
  comparing a large sample of simulated AGN-heated clusters with observations,
  our new simulation technique should make it possible to reliably calibrate
  observational biases in cluster surveys, thereby enabling various
  high-precision cosmological studies of the dark matter and dark energy
  content of the universe.
\end{abstract}

\keywords{methods: numerical, galaxies: clusters: general, cosmology: theory,
  black hole physics}

\section{Introduction}

Numerous observational and theoretical studies of galaxy clusters and groups
show that the astrophysical processes relevant for determining their
properties are still poorly understood. Hydrodynamical simulations of 
cluster formation which only include radiative cooling processes
\citep[e.g.][]{Lewis2000, Muanwong2001, Yoshida2002} suffer from excessive
overcooling within the densest cluster regions, where gas cooling times are
short. This translates into a very large fraction of cold gas and consequently
a large amount of stars that would form out of it, reaching at least $40-50\%$
of the baryonic cluster content. While the associated removal of low-entropy
gas from cluster centers breaks the self-similarity of the cluster scaling
relations in a way that resembles observations
\citep{Bryan2000, Dave2002, Nagai2007}, this `cooling-only' scenario is
generally discounted because observed clusters do not show such strong cooling
flows and enormously bright central galaxies. Instead, it is widely believed
that some non-gravitational energy input must strongly affect the
physics of the intracluster medium (ICM).

There has been considerable effort \citep[see][and references
therein]{Borgani2004} to include feedback mechanisms associated with star
formation in hydrodynamical simulations in order to reduce excessive
overcooling in cluster cores. However, so far simulation
models have failed to simultaneously reproduce the masses and colors of
central galaxies, the observed temperature and metallicity profiles, as well
as the observed X--ray luminosity-temperature ($L_{\rm X}-T$) relation. In
particular, the X--ray luminosities have been found to be substantially larger
than the observed values \citep{Borgani2004} on the group
scale. Furthermore the observed baryon fractions in clusters and
groups are typically smaller than in simulations, which for `standard' physics invariably predict a value close to the universal cosmic baryon fraction \citep[e.g.][]{Frenk1999,OShea2005,Kravtsov2005}.

Several attempts have been made to tackle these
problems with so-called `preheating' schemes
\citep[e.g.][]{Bialek2001,Borgani2005}, that are meant to mimic more energetic
astrophysical processes than the relatively inefficient supernovae feedback.
However, the preheating scenarios are physically not well motivated and rely
on an ad-hoc choice of the amount and epoch of energy injection into the ICM.

On the other hand, it is becoming increasingly clear that active galactic
nuclei (AGN) play an important role for understanding the properties of 
clusters and groups \citep[see ][for a recent
review]{McNamara2007}. Also analytic work
\citep[e.g.][]{Valageas1999, Bower2001, Cavaliere2002} suggests that including
self-consistent feedback from AGN in simulations might resolve the
discrepancies by providing a heating mechanism that offsets cooling, lowers
star formation rates and removes gas from the centers of poor clusters and
groups, thereby reducing their X--ray luminosities to values compatible with
the observed $L_{\rm X}-T$ relation.

In this work we investigate whether AGN feedback can indeed solve these
problems by performing hydrodynamical simulations of galaxy clusters and
groups that employ a state-of-the-art model \citep{Sijacki2007} for BH growth and associated feedback processes.

\section{Methodology} \label{SecMethods}

We have selected a large sample of cluster- and group-sized halos from the
Millennium simulation \citep{Springel2005a} and resimulated them at higher
resolution, including a gaseous component and accounting for hydrodynamics,
radiative cooling, heating by a UV background, star formation and supernovae
feedback. For each halo, two kinds of resimulations were performed. One
containing the physics just described and an additional one including a model
for BH growth and associated feedback processes as in \cite{Springel2005b} and
\cite{Sijacki2007}. This allows us to compare the very same clusters simulated
with and without AGN heating in order to clearly pin down the imprints of AGN
activity on galaxy cluster and group properties, and to which extent this
resolves discrepancies with X--ray observations.

\subsection{The simulations}

In total we have selected $21$ Millennium run dark matter (DM) halos at
$z=0$, and resimulated them at much higher mass and force resolution. The selection was only based on mass and otherwise random, aiming to
approximately uniformly cover a large mass range from $8\!\times\! 10^{12} M_\odot$
to $1.5 \! \times \! 10^{15} M_\odot$. New initial conditions were created by
populating the Lagrangian region of each halo in the original initial
conditions with more particles and adding additional small-scale power, as
appropriate. At the same time, the resolution has been progressively reduced
in regions that are sufficiently distant from the forming halo. Gas
has been introduced into the high-resolution region by splitting
each parent particle into a gas and a DM particle.

We have adopted the same flat $\Lambda$CDM
cosmology as in the parent Millennium simulation, namely $\Omega_{\rm m}=0.25$, $\Omega_{\rm \Lambda}=0.75$, $h=0.73$,
$n_s=1$ and $\sigma_8=0.9$. $\Omega_{\rm b}=0.04136$ has been chosen so as to
reproduce the cosmic baryon fraction inferred from current cosmological
constraints \citep{Komatsu2008}.

In our high-resolution resimulations of halos with virial masses below $2
\! \times \! 10^{14} h^{-1} M_\odot$, the DM particle mass is $m_{\rm DM} =
3.1 \! \times \! 10^7 h^{-1} M_\odot$, the gas particle mass is $m_{\rm gas} = 6.2
\! \times \! 10^6 h^{-1} M_\odot$ and the gravitational softening is $\epsilon = 15
\, h^{-1} \rm kpc$ comoving (Plummer-equivalent) for redshift $z>5$, which is
then replaced with a physical softening of $2.5 \, h^{-1} \rm kpc$ for
$z<5$. For four of these halos we have additionally performed very high resolution simulations
with mass resolution improved by a factor $(3/4)^3$. The change in halo properties resulting from this resolution increase is negligibly small, especially compared to the impact of the AGN heating, indicating that results are approximately converged and our comparison is not affected by resolution issues.
For the four most massive clusters we used a somewhat lower
resolution with a DM particle mass of $m_{\rm DM} = 1.1 \! \times \! 10^8 h^{-1} M_\odot$ and a gas particle mass of $m_{\rm gas} = 2.1
\! \times \! 10^7 h^{-1} M_\odot$. This makes these simulations computationally
affordable, while assuring reasonable convergence.

The simulations were run with the {\small GADGET-3} code \citep[based
on][]{Springel2005c}, which employs an entropy-conserving formulation of
smoothed particle hydrodynamics (SPH). Radiative cooling
and heating was calculated for an optically thin plasma of hydrogen and helium,
and for a time-varying but spatially uniform UV background. We did not include metal-cooling as we wanted to study the impact of AGN feedback independently from the problems involved in following the metal-enrichment of the ICM. One should keep in mind, however, that metal-cooling might have some effect on halo properties on the group scale.
Star formation and supernovae feedback were modeled with a subresolution multi-phase model
for the interstellar medium as in \cite{Springel2003}.

\subsection{The black hole growth and feedback model}

Here we summarize the main features of our model for incorporating BH
growth and feedback in simulations; full details are given in
\cite{Springel2005b}, \cite{Sijacki2006}, and \cite{Sijacki2007}. In short, we assume that
low-mass seed BHs are produced sufficiently frequently that any halo above a
certain threshold mass contains one such BH at its center. In the simulations,
an on-the-fly friends-of-friends group finder puts seed BHs with a mass of
$10^5 h^{-1} M_\odot$ into halos when they exceed a mass of $5
\! \times \! 10^{10} h^{-1} M_\odot$ and do not contain any BH yet. The BHs are
represented by collisionless sink particles and are allowed to grow by mergers
with other BHs and by accretion of gas at the Bondi-Hoyle-Lyttleton rate, but with the Eddington limit
additionally imposed. Two BHs are merged when they fall within their
local SPH smoothing lengths and have small relative velocities.

Motivated by growing theoretical and observational evidence that AGN feedback
is composed of two modes \citep[e.g.][and references therein]{Chrurazov2005}, we use two
distinct feedback models depending on the BH accretion rate (BHAR) itself
\citep[see][]{Sijacki2007}. For large accretion rates above $0.01$ of the Eddington rate, the bulk of AGN heating is
assumed to be in the form of radiatively efficient quasar activity with only a
small fraction of the luminosity being thermally coupled to the ICM. We adopt this thermal heating efficiency to be
$0.5\%$ of the rest mass-energy of the accreted gas, which reproduces the
observed relation between BH mass and bulge stellar velocity dispersion
\citep{DiMatteo2005}. For BHARs below $0.01$ of the Eddington rate, we assume
that feedback is in a so-called ``radio-mode'', where AGN jets inflate hot,
buoyantly rising bubbles in the surrounding ICM. The duty cycle
of bubble injection, energy content of the bubbles as well as their initial
size are determined from the BHAR. We assume the mechanical
feedback efficiency provided by the bubbles to be $2\%$ of the
accreted rest mass-energy, which is in good agreement with observations of
X--ray luminous elliptical galaxies \citep{Allen2006}.

It was shown in \cite{Sijacki2007} that this model leads to a self-regulated BH growth and
brings BH and stellar mass densities into broad agreement with observational
constraints.

\subsection{X--ray properties}
\label{sec:X-ray properties}

We obtain realistic X--ray luminosities and spectroscopic
temperatures for each simulated halo by sorting the gas particles inside
$r_{500}$, the radius that encloses a mean density 500 times the critical
density today, into temperature bins and summing up the emission measure for each
bin. Using the XSPEC package \citep{Arnaud1996}, we then simulate a spectrum of the X-ray emission in that region as a sum of MEKAL emission models \citep{Liedahl1995}, one for each temperature bin, assuming a constant metallicity of 0.3 times the solar value
and using {\em Chandra}'s response function. In order not to be limited by photon
noise we adopted a large exposure time of $10^6 \textrm{s}$. The combined
spectrum of the emission inside $r_{500}$ is then fit by a single temperature
MEKAL model with temperature, metallicity and normalization of the spectrum as
free parameters. The resulting emission model then yields an estimate of the
bolometric luminosity $L_{500}$, while the spectroscopic temperature $T_{500}$
is taken directly from the fitted model. Note, however,
that we use the gas particles inside the three-dimensional radius $r_{500}$ to
calculate $L_{500}$, while we use the gas particles inside the
projected radius $r_{500}$ to obtain $T_{500}$. Also note
that we exclude very cold high-density gas particles with
$T \! < 3 \!\times\! 10^4 \rm K$ and densities above 500 times the mean baryon density as well as multiphase particles to avoid spurious contributions from the multiphase model for the star-forming gas.

\pagebreak

\section{Results} \label{SecResults}

We focus in this work on how AGN feedback affects the $L_{\rm X}-T$ relation and the gas mass fractions in
clusters. Additional cluster properties and simulations where the feedback energy in the ``radio mode'' is not injected
thermally but in the form of cosmic rays \citep{Sijacki2008} will be
discussed in a forthcoming companion paper.

\subsection{Halo gas fractions}

\begin{figure}
 \scalebox{0.72}{\includegraphics{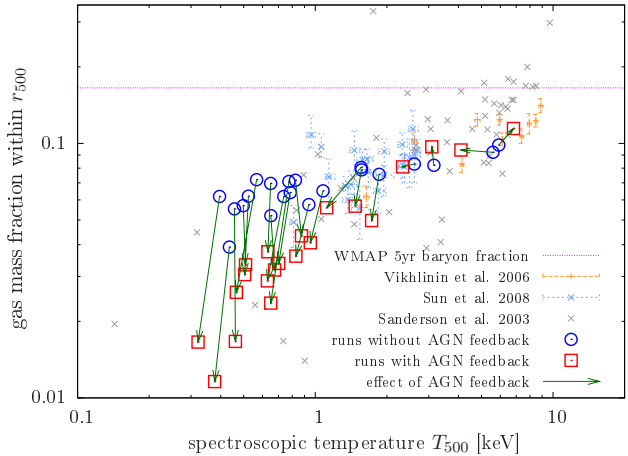}}
\caption{Halo gas mass fractions within $r_{500}$ of clusters
    and groups simulated without AGN feedback (circles) and
    with the feedback model included (squares). The arrows illustrate the
    effect of the AGN heating for each halo.}
\label{fig:gas_fractions}
\end{figure}

In Figure~\ref{fig:gas_fractions}, we show the ratio of
gas mass to total mass inside the radius $r_{500}$ as a function of halo
X--ray temperature. For each simulated halo, arrows connect the
results obtained without and with AGN heating. For
comparison, we show constraints on halo gas fractions obtained from X--ray
observations \citep{Vikhlinin2006,Sun2008}. Also shown are gas fractions we
computed from the gas density and temperature profile parameters given in
\cite{Sanderson2003}.

The most obvious effect of the AGN feedback is the significantly reduced gas
fraction at the low temperature end of our sample, i.e. in poor clusters and
groups. There the AGN heating drives a substantial fraction of the gas to
radii outside of $r_{500}$. This lowers halo gas fractions in spite of the
reduced fraction of gas that is converted into stars in the runs with
AGN. The potential wells of massive clusters are, on the other hand,
too deep for AGN heating to efficiently remove gas from them. Thus the effect
of the suppressed star formation becomes more important or even dominant
towards more massive systems. While the gas fraction in the very inner regions
of massive clusters is somewhat reduced by the AGN, we find it
unchanged or slightly increased within $r_{500}$.

\subsection{The $L_{\rm X}-T$ relation}

\begin{figure*}
  \plotone{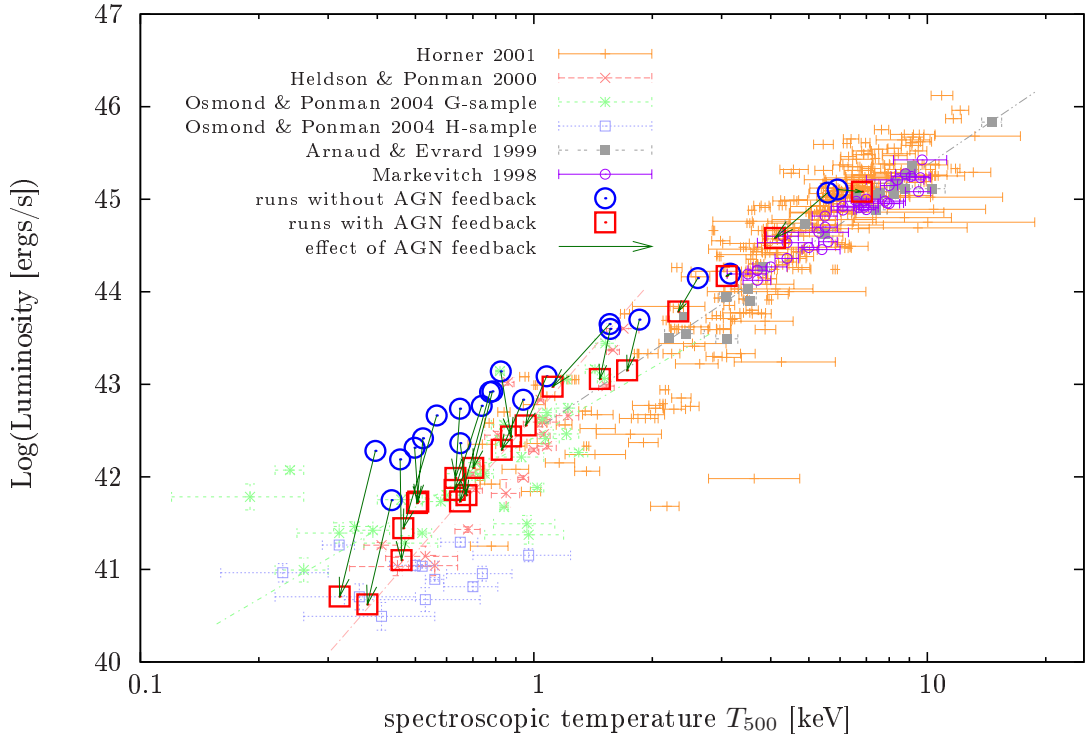}%
\caption{X--ray luminosity as a function of spectroscopic temperature for
  clusters and groups simulated without AGN
  feedback (circles) and with the feedback model included (squares). 
  For each halo, the arrow illustrates the effect of the AGN
  heating. Data from several observational X-ray studies is shown for comparison. For a subset also the best-fit power-law
  $L_{\rm X}-T$ relation is plotted. Including AGN feedback into the
  simulations drastically improves the agreement with observations and
  resolves the discrepancy that otherwise exists on the group scale.}
\label{fig:L_X-T_relation}
\end{figure*}

Given that AGN heating removes gas from the centers of poor clusters and
groups, it is no surprise that it also suppresses their X--ray luminosities
and affects the $L_{\rm X}-T$ relation. In Figure~\ref{fig:L_X-T_relation}, we
plot the X--ray luminosities $L_{500}$ against the spectroscopic temperatures
$T_{500}$, for all halos of our
sample. The arrows indicate the change due to the AGN feedback for each
halo. Data from a number of observational X--ray studies is shown for
comparison
\citep{Horner2001, Helsdon2000, Osmond2004, Arnaud1999, Markevitch1998}.

Without the AGN feedback, we obtain substantially larger X--ray luminosities
for poor clusters than observed, while for massive clusters there
is reasonable agreement. This finding is in line with previous numerical
studies \citep[see e.g.][]{Borgani2004}, and is a manifestation of the
long-standing problem to explain the scaling relations of galaxy clusters
in hydrodynamical simulations. However, when we employ the model for BH growth and feedback the discrepancies between simulated and observed
$L_X-T$ relation are resolved. In particular, the $L_X-T$ relation on the
group scale is steepened significantly, as AGN heating removes a larger
fraction of gas from smaller halos and thereby reduces their X--ray luminosity. For massive clusters, the effect of the
feedback is less important. Overall, the $L_X-T$ relation obtained from the
simulations with AGN feedback is consistent with observations at all mass
scales, from massive clusters to small groups.

Note that while the $L_X-T$ relation of massive clusters is in a
reasonable agreement with observations even without the AGN feedback, this is
only because an unrealistically large amount of cold gas, which is eventually converted into stars, is produced so that the stellar
fractions within the virial radius reach of order of $35-45\%$, in clear
conflict with observations. On the other hand, AGN-heated massive clusters 
not only lie closer to the observed $L_X-T$ relation but also
have much lower stellar fractions which are reduced by at least one
third. Especially their central galaxy is prevented from becoming too bright,
due to the suppression of strong cooling flows. Also, the mass fraction of stars bound to cluster galaxies agrees very well
with observations in our simulations with AGN. There is also a significant component of intra-cluster stars of up to 50\% which
lies at the upper end of the observational estimates \citep[see e.g.][for an overview]{Lin2004}.

\section{Summary and conclusions}  \label{SecConclusions}

We have performed very high-resolution numerical simulations of a
mass-selected sample of galaxy clusters and groups. In order to investigate
whether AGN feedback makes simulations and X--ray observations of cluster
and groups compatible, we have carried out two kinds of simulations for each
halo, namely (1) hydrodynamical simulations with a treatment of radiative
cooling, star formation, supernovae feedback, and heating by a UV background,
and (2) simulations that additionally employ a model for black hole growth and
associated feedback processes. Our main findings are:
\begin{itemize}
\item AGN feedback significantly lowers the gas mass fractions in poor
  clusters and groups, even though fewer baryons are turned into stars at the
  same time. This is because the AGN heating drives gas from
  halo centers to their outskirts. In massive clusters, on the other hand, it mainly lowers the central gas density and
  substantially reduces the amount of stars formed. Overall, both the gas and
  stellar fractions of the whole sample of our simulated groups and clusters
  are in a much better agreement with observations when AGN are included
  than in simulations without AGN.
\item AGN feedback significantly reduces the X--ray luminosities of poor
  clusters and groups, while the X--ray temperature within $r_{\rm 500}$ stays
  roughly the same or is even slightly reduced. This results in a steepening
  of the $L_{\rm X}-T$ relation on the group scale.
\item The $L_{\rm X}-T$ relation obtained from simulations with AGN feedback
  is in excellent agreement with observations at all mass scales.
\end{itemize}

We find it extremely encouraging that this simple model for BH growth and
feedback is capable of bringing the analyzed
properties of galaxy clusters and groups into much better agreement with
observations. This not only resolves a
long-standing problem in their hydrodynamical modeling, but
also opens up new exciting possibilities for using numerical simulations to
investigate the properties of clusters and groups, and their
co-evolution with central supermassive BHs. Also, it considerably brightens
the prospects to use simulations of cluster formation to accurately
calibrate and correct systematic effects in future X-ray and
Sunyaev-Zel'dovich cluster surveys. This is essential for the exploitation of
the full potential of clusters as cosmological probes to accurately constrain
the expansion history of the Universe.

\acknowledgements We thank Simon White, Martin Haehnelt, Klaus Dolag and Gabriel Pratt for constructive discussions. D.S. acknowledges Postdoctoral Fellowship from the STFC. Part of the simulations were run on the Cambridge HPC cluster Darwin.

\bibliographystyle{apj}

\end{document}